\documentclass[
aps,
nofootinbib,
superscriptaddress,
tightenlines,
notitlepage,
twocolumn,
showpacs,
floatfix,
]{revtex4-1}
\usepackage{amssymb,amsmath,bm,tensor,braket}
\usepackage[colorlinks]{hyperref}

\usepackage{graphicx}
\usepackage{dcolumn}
\usepackage{bm}
\usepackage{xcolor}

\newcommand{\BIT}{\affiliation{School of Physics, Beijing Institute of Technology, Beijing, 100081, China}}

\begin{document}

\preprint{APS/123-QED}

\title{Entanglement Islands and Thermodynamics of the Black Hole \\ in Asymptotically Safe Quantum Gravity}

\author{Sobhan Kazempour}\email{sobhan.kazempour1989@gmail.com}\BIT
\author{Sichun Sun}\email{sichunssun@gmail.com}\BIT
\author{Chengye Yu}\email{chengyeyu1@hotmail.com}\BIT

\begin{abstract}
We study thermodynamic properties and the entanglement island of a black hole in asymptotically safe quantum gravity, analyzing key thermodynamic quantities such as the Hawking temperature, heat capacity, and entropy, as well as the mass-horizon radius relation. Unlike Schwarzschild black holes, the temperature decreases with mass near the evaporation endpoint, signaling a phase transition and possible stable remnant. The entanglement entropy of Hawking radiation is obtained both with and without island contributions. Without islands, the radiation entropy grows linearly indefinitely, leading to the information paradox. By including island contributions and extremizing the generalized entropy functional, we resolve this paradox. At late times, the radiation entropy saturates at the Bekenstein-Hawking entropy, confirming unitary evolution. From this, we derive the Page time and scrambling time by equating early- and late-time entanglement entropies. The result of this study establishes the finiteness of the radiation entropy and consistency with quantum mechanics.
\end{abstract}


\maketitle


\section{Introduction}\label{sec:1}
Black holes are strange objects predicted by the Einstein field equations and have fascinated scientists since Albert Einstein first worked on general relativity \cite{Einstein:1916vd,Schwarzschild:1916ae,Reissner:1916cle,Weyl:1917rtf,Kerr:1963ud,Stephani:2003tm,Banados:1992wn,Myers:1986un,Garfinkle:1990qj}. While the no-hair theorem predicts that black holes can be specified entirely, or defined by only three parameters, mass, electric charge, and angular momentum, there are other claims too \cite{Israel:1967wq,Israel:1967za,Carter:1971zc}.
The study of black hole thermodynamics emerged in the 1970s \cite{Bardeen:1973gs,Bekenstein:1972tm,Bekenstein:1973ur,Bekenstein:1974ax,Bekenstein:1975tw,Hawking:1975vcx} and has provided further insight into these astronomical bodies. Black holes are ideal theoretical laboratories for studying the nature of gravity, particularly in strong-field regimes. The relationship between black holes and classical thermodynamics, formulated by Bekenstein in 1973 \cite{Bekenstein:1973ur}, has led to a deeper understanding of gravitational physics and thermodynamics.

The thermodynamic context also reveals several interesting features. The temperature of a Schwarzschild black hole is inversely proportional to its mass, and it has negative heat capacity \cite{Bekenstein:1972tm,Bardeen:1973gs,Hawking:1974rv}. Negative heat capacity signifies inherent thermodynamic instability \cite{Davies:1978zz}. A Reissner-Nordström black hole or a Kerr-Newman black hole has regions of positive and negative heat capacity, and more complicated stability properties \cite{Davies:1978zz,Sokolowski:1980uva,Cai:1997cs}. Many studies have examined various features of thermodynamic properties with respect to different theories of gravity and black hole solutions \cite{Padmanabhan:2009vy,Kubiznak:2016qmn,Hale:2025veb,Wei:2022dzw,Kruglov:2022mde,Frassino:2022zaz,Cassani:2022lrk,Kapec:2023ruw,Misyura:2024fho,Elizalde:2025iku}.

Black hole thermodynamics has been studied extensively, and one can assign entropy to the horizon area \cite{Bardeen:1973gs} and temperature to the surface gravity \cite{Hawking:1975vcx}. We can note here on the Hawking-Page phase transition \cite{Hawking:1982dh}, analogous to confinement/deconfinement transitions for a conformal field theory (CFT) \cite{Maldacena:1997re,Witten:1998qj}. The advent of the anti-de Sitter (AdS)/CFT correspondence \cite{Maldacena:1997re,Maldacena:1998zhr,Aharony:1999ti,Shen:2005nu} provided a new lens through which to study black hole thermodynamics. Stephen Hawking,  Gerard’t Hooft, and others proposed the holographic principle and suggest that information is written on the event horizon \cite{Hawking:1976ra,tHooft:1993dmi}.

It was later shown that the thermodynamics of AdS space can be dual to a CFT on its boundary \cite{Witten:1998qj}, and that this duality may unify general relativity and quantum mechanics via theories of quantum gravity such as string theory or M-theory. The black hole information paradox lies at the intersection of quantum mechanics, thermodynamics, and general relativity, and remains an outstanding puzzle \cite{Hawking:1975vcx,Page:1993wv,Wald:1975kc,Parker:1975jm}. Recent developments have given some glimpses of what it might mean for information to be preserved during black hole evaporation \cite{Almheiri:2019hni,Almheiri:2019yqk,Almheiri:2019qdq,Almheiri:2019psf}, and provide a quantum description without a complete understanding of the quantum dynamics of the black hole, and without a complete theory of quantum gravity.

According to Stephen Hawking's 1975 work, black holes emit particles via a process called Hawking radiation \cite{Hawking:1975vcx}. Over time, this radiation will cause black holes to lose mass and eventually evaporate. However, Hawking's calculation revealed a more serious problem; the radiation is just a thermal state without quantum correlations. This apparent incompatibility leads to a paradox when applying the unitarity principle to a black hole formed from an initial pure quantum state, since unitarity holds only for pure states evolved deterministically \cite{Hawking:1976ra}.

The paradox also manifests in terms of the evolution of entropy. According to Hawking, the von Neumann entropy of the radiation diverges during the evaporation process; however, it is in direct conflict with the quantum-mechanical limit on entropy for a finite quantum system. If the evaporation process were unitary, however, the entropy in the radiation must first widen to reach the Bekenstein-Hawking entropy (thermodynamic quantity related to the area of the black hole horizon \cite{Bekenstein:1980jp}) and then fall quickly back to zero, as the black hole evaporates. The expected trajectory of entropy, which increases and then decreases, is known as the Page curve \cite{Page:1993wv,Page:2013dx}, and it is consistent with unitary evolution. The resolution of the information paradox hinges on deriving the Page curve from a physical context.

Page presented that a quantum system, such as a black hole, evolving unitarily from an initial pure state, will have entanglement entropy that increases linearly with time during the initial evaporation period and will decrease to zero once the black hole has radiated all its energy \cite{Page:1979tc,Giddings:1992fp} largely. Several proposals have been put forward to resolve the information paradox, each with its own merits and drawbacks \cite{Hawking:1982dj,Bekenstein:1980jp,Harlow:2014yka}. Proposals include backreaction to arrive at a final pure state, information emission at the end of black hole evaporation at the Planck scale, a Planck-scale remnant, baby universes as mechanisms of information dissipation, and an entirely new process of information emission. Unfortunately, some of these proposals encountered problems, in the form of violations of the Bekenstein bound \cite{Banks:1983by}, the need for causal information to escape \cite{Giddings:1988cx}, or violation of causality at the horizon \cite{Giddings:1988cx,Giddings:1992hh}. Hence, Parikh and Wilczek suggested that the information paradox could be studied by considering a higher-order non-thermal effect in the radiation, to allow otherwise, information to leak out from the black hole \cite{Parikh:1999mf}; however, this effect is negligible with respect to massive black holes and does not make up for all the information lost in such circumstances.

The question of whether black hole dynamics violate unitarity remains open. The AdS/CFT correspondence \cite{Maldacena:1997re}, the most essential mathematical incarnation of black hole complementarity \cite{Susskind:1993if}, has made the most progress. It may be possible because information is conserved when the black hole in AdS space is mapped onto a unitary CFT at its boundary, which would imply that evaporation should proceed through the Page curve. However, it was only recently that the Page curve could be derived, and a "firewall" must exist at the horizon for evaporation to be unitary \cite{Almheiri:2012rt}, violating "no drama" in general relativity. Consequently, achieving a definitive solution to the black hole information paradox remains an essential research priority.

The recent advances in our understanding of Hawking radiation and the black hole information paradox arose from the calculation of the Page curve for two-dimensional black holes in AdS spacetimes in the context of Jackiw-Teitelboim (JT) gravity semi-classically \cite{Almheiri:2019yqk, Almheiri:2019qdq}. The focus of research has been primarily on models of two-dimensional gravity, due to their uniqueness in guaranteeing the existence of exact solutions, while accounting for the backreaction of the radiation when using semi-classical approximations \cite{Rozali:2019day,Chen:2019uhq,Chen:2020jvn,Gautason:2020tmk,Chen:2020wiq,Almheiri:2020cfm}. One promising development is the island paradigm \cite{Almheiri:2019hni,Penington:2019npb,Almheiri:2019psf}, which reproduces the Page curve and reconciles black hole evaporation with unitarity by the regions of spacetime called islands in the calculations of entropy. The protocol uses a generalized entropy \cite{Bekenstein:1973ur}, which combines the Bekenstein-Hawking entropy with contributions from quantum fields. Islands are bounded by quantum extremal surfaces (QES) \cite{Engelhardt:2014gca}, which are extremal points in the geometries of the spacetime. To compute the fine-grained entropy of the radiation, you minimize the generalized entropy functional to find the possible island configurations, and the entropic value is then the minimum value. An important point is that this value for entropy is rigorously obtained from the gravitational path integral \cite{Penington:2019kki, Almheiri:2019qdq}, and as such, it has a solid foundation in semi-classical gravity.

Island geometry is crucial for determining generalized entropy for four-dimensional asymptotically flat black holes. Prior work also has focused on static configurations, specifically Schwarzschild \cite{Hashimoto:2020cas,Arefeva:2021kfx} and Reissner Nordstrom (RN) black holes \cite{Kim:2021gzd,Wang:2021woy,Matsuo:2021mmi}. Meanwhile, studies have focused on dynamic systems, with initial descriptions of one-sided dynamical black holes \cite{Alishahiha:2020qza} invoking the eternal black hole-centered Hartle-Hawking (HH) state \cite{Hartle:1976tp} in a controversial manner. To reconcile this problem, the "in" vacuum state \cite{Fabbri:2005mw} was introduced for dynamical black holes obtained from collapsing null shells \cite{Gan:2022jay}. This had been successfully generalized to evaporating black holes using the Vaidya metric \cite{Guo:2023gfa}, proving that the entire conceptual framework can be generalized in time-dependent, gravitationally dynamic cases. Finally, extensive research and discussion have gone into analyzing entanglement islands in various black hole solutions \cite{Cadoni:2021ypx,Hollowood:2020cou,Liu:2020gnp,Lin:2024gip,Luongo:2023jyz,Almheiri:2019psy,Hartman:2020swn,Guo:2023gfa,Uhlemann:2021nhu}.

We now focus on the black hole in asymptotically safe quantum gravity. The study investigates the framework of asymptotically safe quantum gravity, which posits that gravity can be consistently defined at all scales due to a nontrivial ultraviolet fixed point that gives predictivity without additional dimensions and/or exotic fields \cite{Reuter:2019byg,Bonanno:2020bil,Reichert:2020mja}. Based on the functional renormalization group, this framework avoids unphysical divergences while maintaining background independence, and therefore is a potential pathway to a unified quantum theory of gravity \cite{Denz:2016qks,Fraaije:2022uhg}. Recently, the black hole solutions of asymptotically safe gravity have been examined for their properties and consequences related to quantum spacetime \cite{Knorr:2019atm,Falls:2014tra,Falls:2018ylp,Falls:2010he,Cai:2010zh,Becker:2012js,Saueressig:2015xua,Bonanno:2023rzk,deBrito:2023ydd,Stashko:2024wuq,Platania:2025imw}. In this study, we will primarily assess thermodynamics aspects and the entanglement island for the black hole in asymptotically safe quantum gravity, and determine whether or not methods from \cite{Gan:2022jay} can be used to resolve the information loss paradox and, ultimately, elucidate the preservation of quantum information in black hole evaporation.

The paper is organized as follows. In Sec. \ref{sec:2}, we discuss the black hole metric within the context of asymptotically safe quantum gravity and analyze its thermodynamic properties. In Sec \ref{sec:3}, we compute the entanglement entropy of Hawking radiation for both the island-free and island-containing scenarios. Meanwhile, we determine the Page time and scrambling time by analyzing the entanglement entropy values at early and late stages of the system's evolution. In the final section \ref{sec:4}, we present our conclusions and summarize our findings.

\section{REVIEW OF THERMODYNAMIC PROPERTIES OF BLACK HOLE IN ASYMPTOTICALLY SAFE QUANTUM GRAVITY}\label{sec:2}

In this section, we introduce the black hole metric within the framework of asymptotically safe quantum gravity. We also review its key features and analyze its thermodynamic properties.

In the context of asymptotically safe quantum gravity, black holes are characterized by solutions that emerge from the quantum effective action and the corresponding quantum equations of motion. These solutions are constructed using multigraviton correlation functions within the asymptotic safety framework and have recently been the subject of numerical investigations \cite{Pawlowski:2023dda}.

In the framework of asymptotically safe quantum gravity, the metric for a static, spherically symmetric spacetime is given by
\begin{eqnarray}\label{ds1}
d s^{2} = -f(r) d t^{2} + g(r)^{-1} d r^{2} + r^{2} (d\theta^{2} + sin^{2} \theta d \phi^{2}),  \nonumber\\
\end{eqnarray}
here, the metric function is described by:
\begin{eqnarray}
f(r) &&= 1 - \frac{2 M}{r} + S_{0} \frac{e^{-m_{0}r}}{r} + S_{2}\frac{e^{-m_{2}r}}{r},\nonumber\\
g(r) &&= 1 - \frac{2 M}{r} - S_{0} \frac{e^{-m_{0}r}}{r} (1+ m_{0}r) \nonumber\\ && + \frac{1}{2} S_{2}\frac{e^{-m_{2}r}}{r} (1+ m_{2}r). \nonumber\\
\end{eqnarray}
Notice that the lapse functions $f(r)$ and $g(r)$, which depend on the radial coordinate $r$. You can get the Schwarzschild metric by making $f(r)=g(r)= 1 - \frac{2 M}{r}$ as long as $(S_{0}, S_{2}) = (0, 0)$ Here, $M$ stands for the mass of the black hole as seen by someone observing at infinity. The two free parameters, $S_{0}$ and $S_{2}$, control how strong the Yukawa corrections are, which decay. Also, the masses $m_{0}$ and $m_{2}$, which correspond to the spin-0 and spin-2 modes, determine how the different curvature terms couple.

In line with the study \cite{Pawlowski:2023dda}, which focused only on the effect of the lightest additional mode originating from the higher curvature operators, this means that it is dependent on the solutions on $S_{0}$, and the $S_{2}$ is fixed. In this paper, we have similarly fixed $S_{2}=1$ and present results for various $S_{0}$.

Furthermore, it should be noted that by considering the $S_{2}=1$ and varying the $S_{0}$, there are three solutions. The first case is related to the negative values of $S_{0}$, which are horizon-less geometries. Moreover, in the positive case of $S_{0}$ the solutions include a radius $r_{h}$ where $g(r_{h})=0$, if the $S_{0}$ would be smaller than a critical value $S_{0,c}$, we have type II solution. Also, we have a type III solution if the $S_{0}$ would be bigger than $S_{0,c}$ \cite{Pawlowski:2023dda}.

\begin{figure*}
\centering
{(a) }\includegraphics[width=0.37\linewidth]{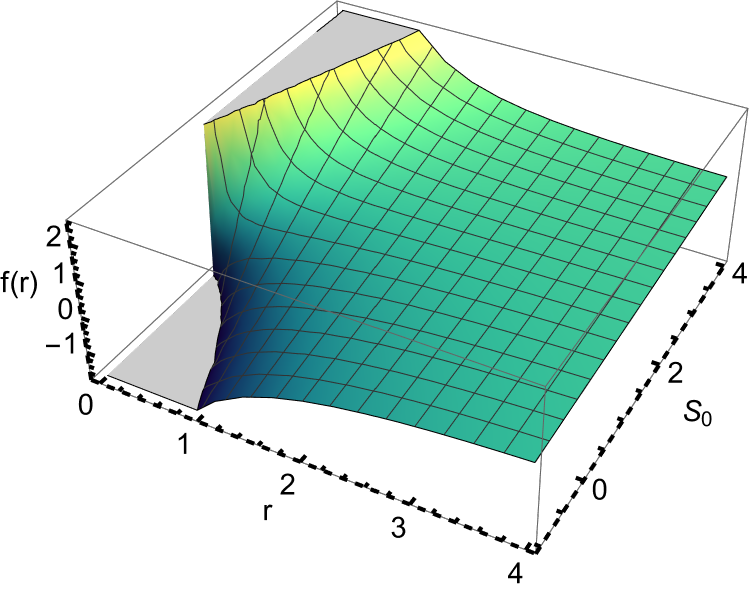}
\includegraphics[width=0.065\linewidth]{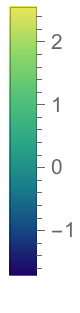}
{(b) }\includegraphics[width=0.26\textwidth]{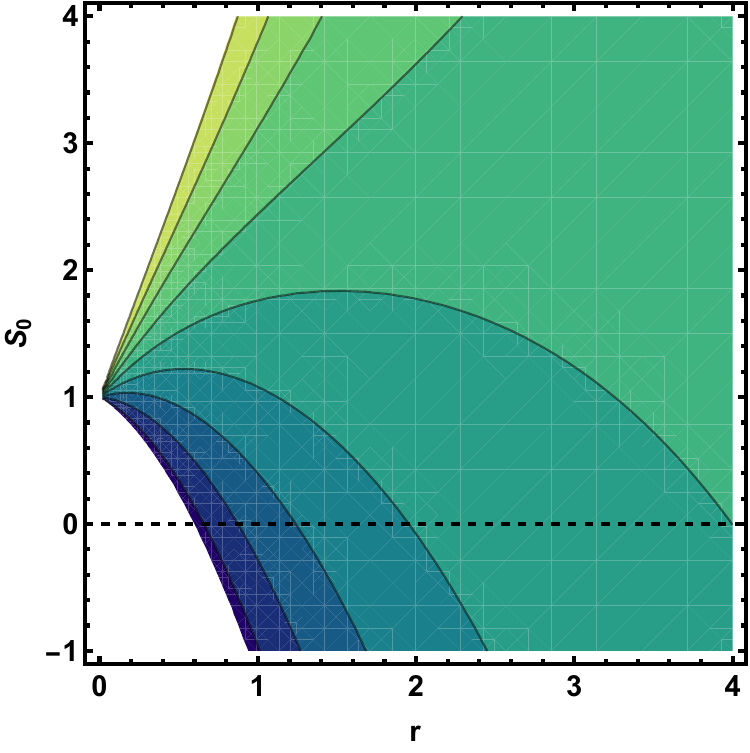}
\includegraphics[width=0.045\linewidth]{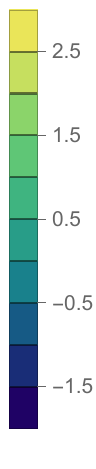}
\caption{The profile of $f(r)$ with respect to $r$ for various of $S_{0}$ by considering $M=1$, $S_{2}=1$, $m_{2}^{2}=2.5 M_{pl}^{2}$ and $m_{0}^{2}=0.095 M_{pl}^{2}$ \cite{Pawlowski:2023dda}.}
\label{fig01}
\end{figure*}

\begin{figure*}
\centering
{(a) }\includegraphics[width=0.37\linewidth]{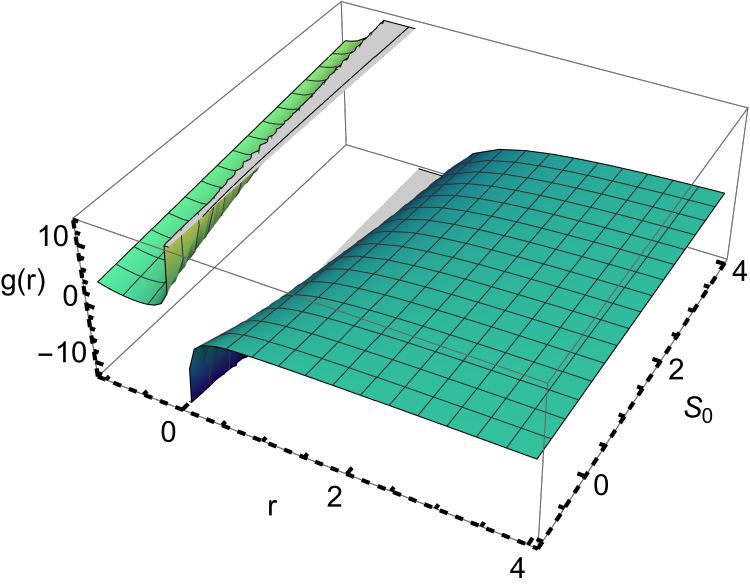}
\includegraphics[width=0.065\linewidth]{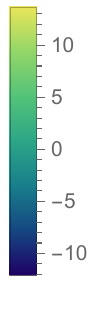}
{(b) }\includegraphics[width=0.26\textwidth]{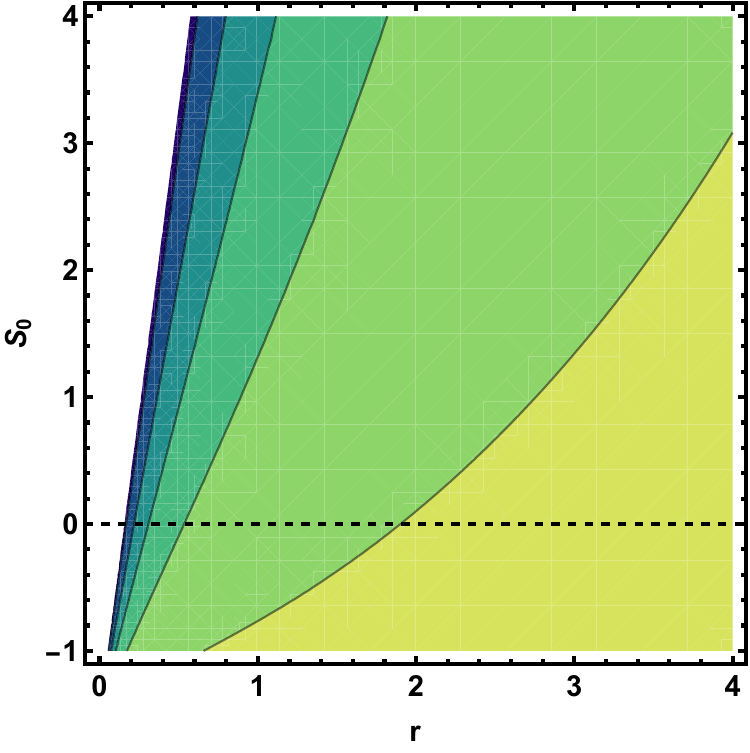}
\includegraphics[width=0.045\linewidth]{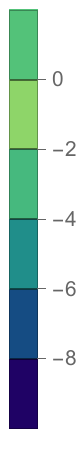}
\caption{The profile of $g(r)$ with respect to $r$ for various of $S_{0}$ by considering $M=1$, $S_{2}=1$, $m_{2}^{2}=2.5 M_{pl}^{2}$ and $m_{0}^{2}=0.095 M_{pl}^{2}$ \cite{Pawlowski:2023dda}.}
\label{fig02}
\end{figure*}

In Figures (\ref{fig01}) and (\ref{fig02}), we show the profile of $f(r)$ and $g(r)$ of the black hole in asymptotically safe quantum gravity. We illustrate the change of the profiles of $f(r)$ and $g(r)$ for various values of $S_{0}$ by assuming determined values for $M$, $S_{2}$, $m_{0}$, and $m_{2}$ \cite{Pawlowski:2023dda}.


Notice that the event horizons are determined by the condition $g^{rr}=g(r_{h}) = 0$. However, in this case, because we have exponential terms (i.e., $e^{-m_{0}r}$ and $e^{-m_{2}r}$), computing exact solutions for event horizons is difficult. So, we consider the approximate approach, and we expand the exponential terms in a series $e^{x}= 1 + \frac{x}{1 !}+\frac{x^{2}}{2!}+...$. By substituting the series expansion for the exponential terms, we can approximately obtain the event horizons as functions of the theory's parameters. The event horizon is obtained,

\begin{eqnarray*}
r_{h}= \frac{1}{3\big(- 2 m_{0}^{3}S_{0} + m_{2}^{3}S_{2} \big)} \Bigg\lbrace    -2 m_{0}^{2}S_{0} +m_{2}^{2}S_{2} + \bigg[ 4 m_{0}^{3}S_{0} (m_{0}S_{0}\nonumber\\ + 6) - 4 m_{2}^{2}(m_{0}^{2}S_{0}  +3 m_{2}) S_{2} + m_{2}^{4}S_{2}^{2} \bigg] \times \Xi^{-\frac{1}{3}} + \Xi^{\frac{1}{3}}  \Bigg\rbrace,  
\end{eqnarray*}
\begin{widetext}
\begin{eqnarray}\label{rh}
\Xi=  \Bigg\lbrace 12 m_{0}^{4}m_{2}^{2}S_{0}^{2}S_{2} -72 m_{0}^{5}S_{0}^{2} - 6 m_{0}^{2}m_{2}^{3}S_{0}S_{2}(m_{2}S_{2}-6) + 4 m_{0}^{6}S_{0}^{2}\big( 52 S_{0} - 27 (S_{2} - 4 M) \big) -36 m_{0}^{3}m_{2}^{2}S_{0}S_{2} \big( -1 \nonumber\\ + 3m_{2} ( 2 S_{0} - S_{2} + 4 M ) \big) + 2 m_{2}^{5} S_{2}^{2} \big( - 9 + m_{2} (27 S_{0} - 13 S_{2} + 54 M) \big) +  \bigg[ \big( 12(-2 m_{0}^{3}S_{0} +m_{2}^{3}S_{2}) - (-2 m_{0}^{2}S_{0} \nonumber\\ + m_{2}^{2}S_{2})^{2} \big)^{3}  + 4 \bigg( -36 m_{0}^{5}S_{0}^{2}  + 6 m_{0}^{4}m_{2}^{2}S_{0}^{2}S_{2}  - 3 m_{0}^{2}m_{2}^{3}S_{0}S_{2} ( m_{2}S_{2} - 6) + 2 m_{0}^{6}S_{0}^{2} \big( 52 S_{0} - 27 ( S_{2} - 4 M ) \big) \nonumber\\ - 18 m_{0}^{3}m_{2}^{2}S_{0}S_{2} \big(- 1 + 3 m_{2} ( 2 S_{0} - S_{2}  +4 M ) \big)  + m_{2}^{5} S_{2}^{2} \big( -9 + m_{2} ( 27 S_{0} - 13 S_{2} +54 M )  \big) \bigg)^{2} \bigg]^{\frac{1}{2}} \Bigg\rbrace.  \nonumber\\
\end{eqnarray}
\end{widetext}

\begin{figure*}
\centering
{(a) }\includegraphics[width=0.37\linewidth]{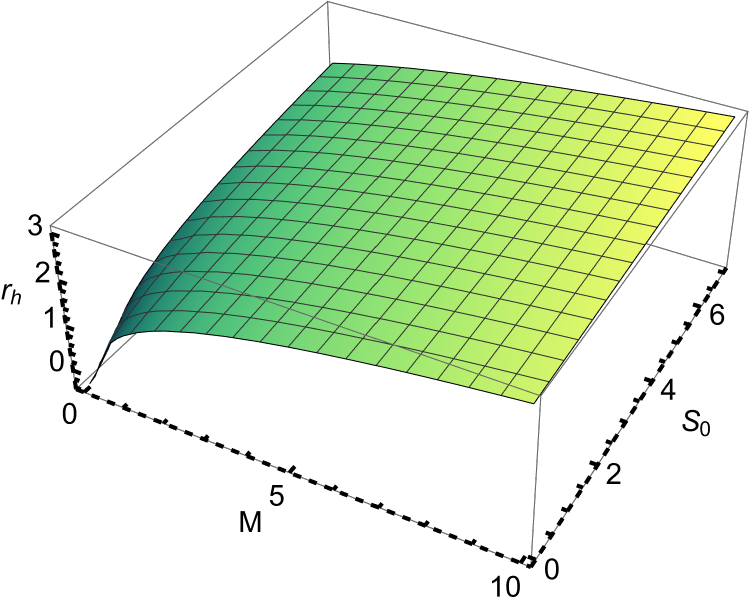}
\includegraphics[width=0.065\linewidth]{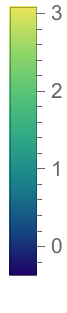}
{(b) }\includegraphics[width=0.26\textwidth]{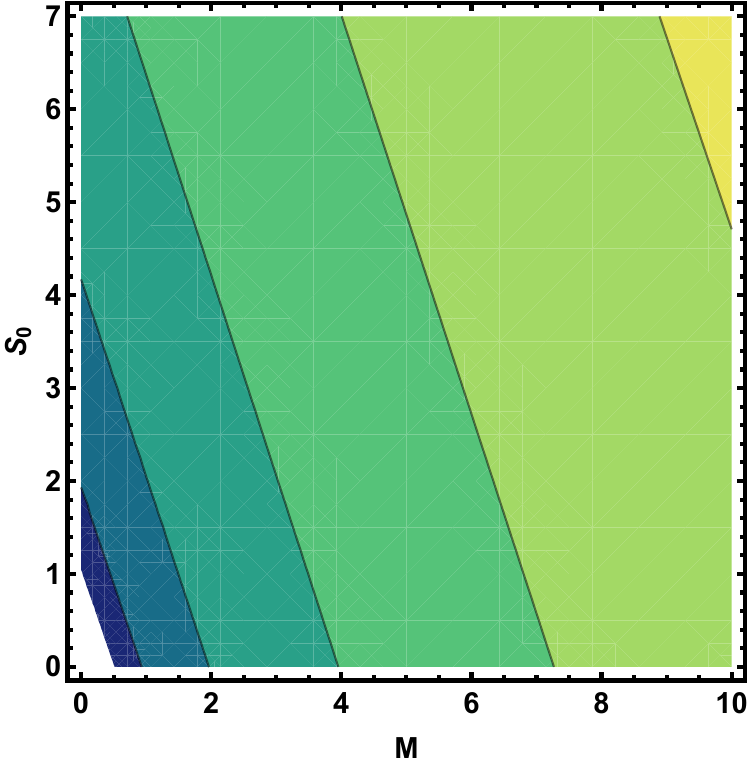}
\includegraphics[width=0.045\linewidth]{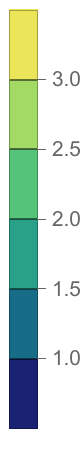}
\caption{The graphs illustrate the possible event horizons $r_{h}$ with respect to variety values of $M$ and $S_{0}$. We used $m_{2}^{2}=2.5 M_{pl}^{2}$ and $m_{0}^{2}=0.095 M_{pl}^{2}$ \cite{Pawlowski:2023dda}.}
\label{fig03}
\end{figure*}
In Figure (\ref{fig03}), we have shown the different possible ranges of event horizons $r_{h}$ for a variety of values of $M$ and $S_{0}$. 
It is clear that as the black hole mass $M$ and the $S_{0}$ parameter increase, the event horizon also increases. 

Furthermore, for $S_{0} < S_{0,c}$, $r_{h}$ scales linearly with $M$ (e.g., $r_{h} \approx 2 M$ for $S_{0}=0$), mimicking Schwarzschild behavior. As $S_{0}$ approaches $S_{0,c}$ from below, $r_{h}$ increases at fixed $M$, indicating horizon expanding. For $S_{0} > S_{0,c}$, solutions transition abruptly to type III black holes with $r_{h}$ discontinuously jumping to a higher branch. A critical mass $M_{min}(S_{0})$ exists below which no horizon forms (horizonless regime).

In the following discussion, we explore the thermodynamic properties of the black hole metric within the framework of asymptotically safe quantum gravity. The Hawking temperature $T=\frac{1}{4\pi}\sqrt{|f^{'}(r_{h})g^{'}(r_{h})|}$ for the metric given by \cite{Pawlowski:2023dda},
\begin{widetext}
\begin{eqnarray}\label{T}
T= \frac{1}{4 \sqrt{2}\pi r_{h}^{2}} \Bigg\lbrace - e^{-2 r_{h}(m_{0}+m_{2})} \bigg( e^{r_{h} m_{2}} (1 + r_{h}m_{0}) S_{0} + e^{r_{h}m_{0}} \big( -2 e^{r_{h} m_{2}} M + (1+ r_{h}m_{2}) S_{2} \big) \bigg) \bigg( 2 e^{r_{h} m_{2}} \big(1 \nonumber\\ +r_{h}m_{0}(1 + r_{h} m_{0}) \big) S_{0} + e^{r_{h}m_{0}} \big( 4 e^{r_{h}m_{2}} M + (1 + r_{h} m_{2} (1 + r_{h}m_{2})) S_{2} \big) \bigg) \Bigg\rbrace^{\frac{1}{2}}. \nonumber\\
\end{eqnarray}
\end{widetext}

\begin{figure*}
\centering
{(a) }\includegraphics[width=0.37\linewidth]{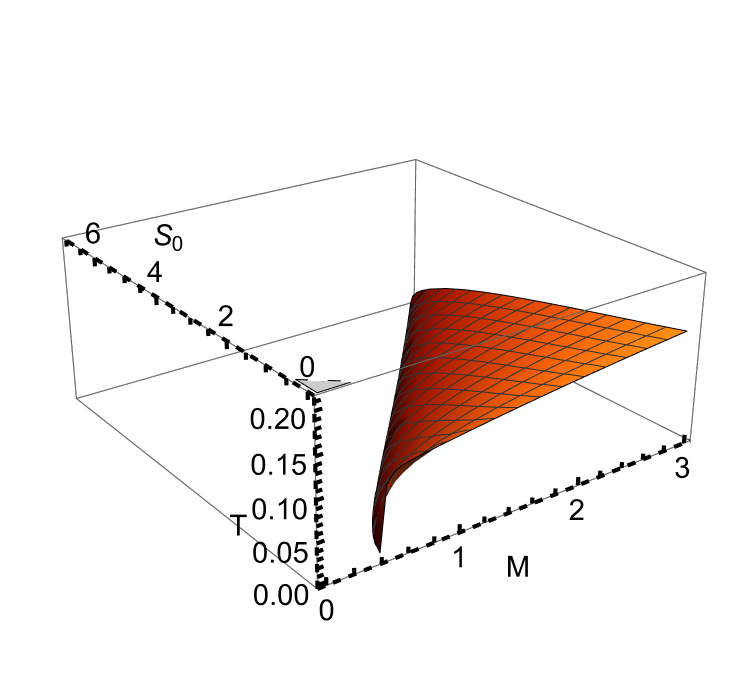}
\includegraphics[width=0.08\linewidth]{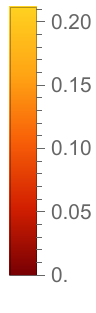}
{(b) }\includegraphics[width=0.26\textwidth]{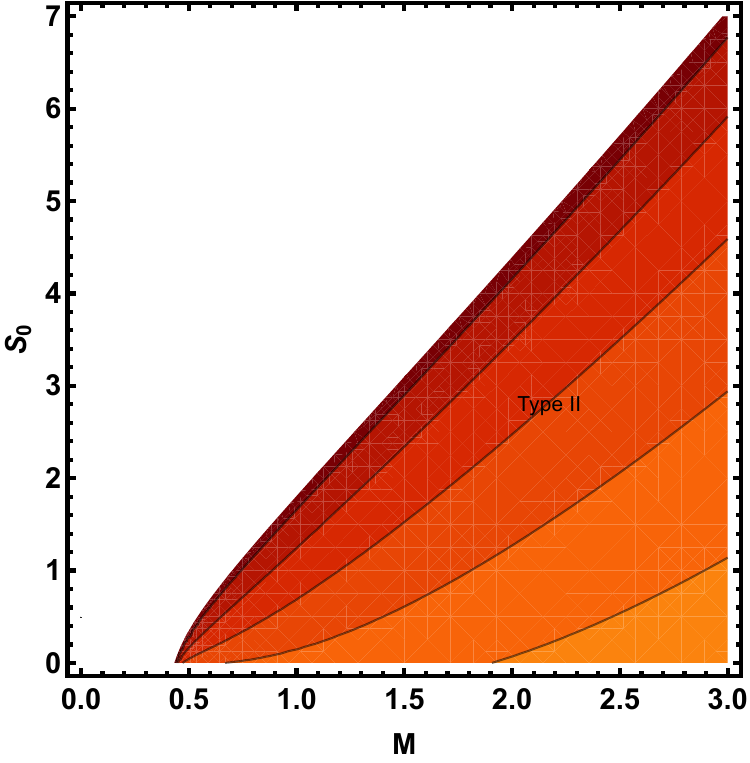}
\includegraphics[width=0.055\linewidth]{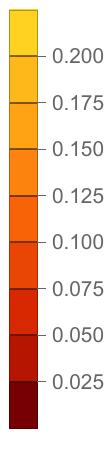}
\caption{The graphs present the possible Hawking temperature $T (M, S_{0})$ with respect to variety values of $M$ and $S_{0}$. We used $m_{2}^{2}=2.5 M_{pl}^{2}$ and $m_{0}^{2}=0.095 M_{pl}^{2}$ \cite{Pawlowski:2023dda}.}
\label{fig04}
\end{figure*}

In Figure (\ref{fig04}), we have illustrated the possible ranges of Hawking temperature $T$ with respect to different values of $M$ and $S_{0}$.
Notice that the other disappeared regions of the graph show the type I and III solutions, which are presented on \cite{Pawlowski:2023dda}. As mentioned earlier, type I is horizonless geometries, and type III is the exotic case. Here, we only focus on the modified black hole solutions of type II.

The observed thermal gradient across this portion of the black hole spectrum is not consistent with classical prediction, where lower mass black holes have higher temperature, with $T_{H}\varpropto \frac{1}{M}$. Significantly, the endpoint of Hawking evaporation raises critical theoretical issues. According to Hawking radiation, these black holes are expected to lose mass gradually and move to the left of the graph as they enter the phase-transition region, where the temperature vanishes and evaporation stops.
The discontinuity in temperature at critical $S_{0}$ reflects a first-order phase transition between a Type II and Type III black hole. Since the discontinuity indicates a thermodynamic instability as evaporation proceeds, the black hole would no longer lose mass continuously and smoothly. Of primary importance, asymptotic safety provides for a minimum mass, $M_{min}(S_{0})$, below which horizons cease to exist, signifying a stable remnant. These remnants are a potential solution to curvature singularities and provide a unitary evolution, so long as they exist, asymptotically safe quantum gravity does not share the same fate as semiclassical gravity. In semiclassical gravity, the end of evaporation is catastrophic.

Subsequently, the relationship between the mass parameter $M$, and the horizon radius $r_{h}$ is derived as follows:

\begin{eqnarray}\label{M}
M =  \frac{1}{4} e^{- r_{h} (m_{0} + m_{2})} \bigg( 2 r_{h} e^{r_{h} (m_{0} + m_{2})} - 2 S_{0} e^{r_{h} m_{2}} \nonumber\\ - 2 r_{h} m_{0} S_{0} e^{r_{h} m_{2}} + S_{2} e^{r_{h}m_{0}} + r_{h} m_{2} S_{2} e^{r_{h} m_{0}}  \bigg).
\end{eqnarray}

\begin{figure*}
\centering
{(a) }\includegraphics[width=0.37\linewidth]{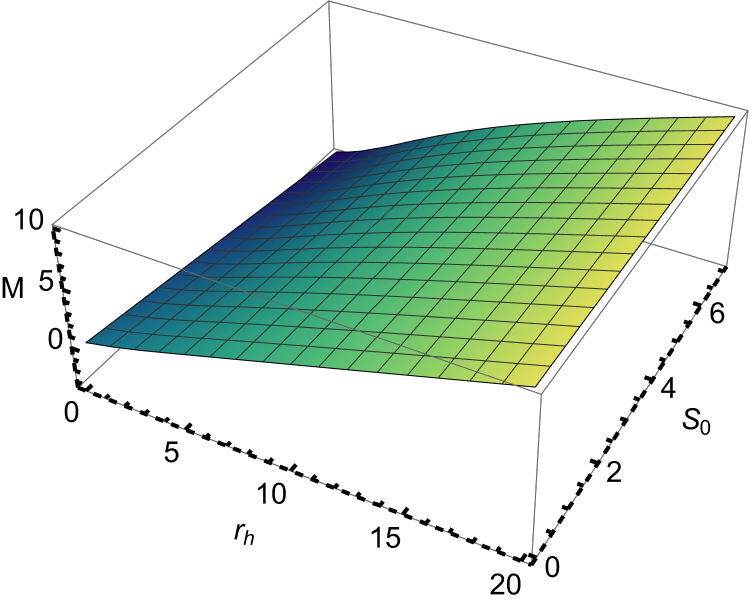}
\includegraphics[width=0.08\linewidth]{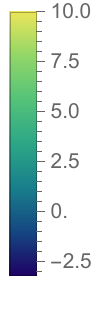}
{(b) }\includegraphics[width=0.26\textwidth]{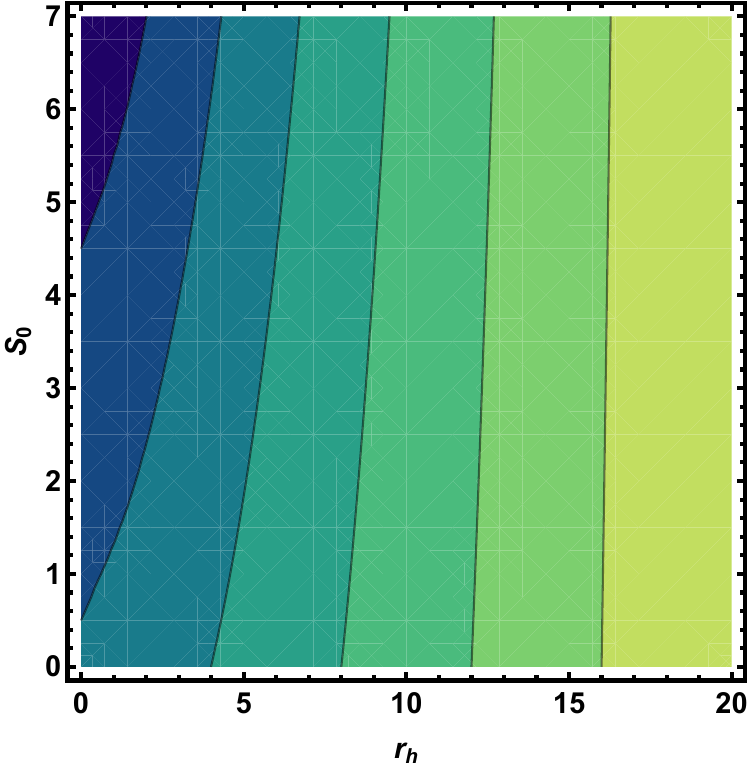}
\includegraphics[width=0.055\linewidth]{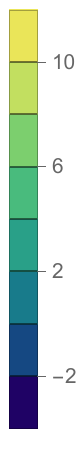}
\caption{The graphs present the possible mass $M$ with respect to variety values of $r_{h}$ and $S_{0}$. We used $m_{2}^{2}=2.5 M_{pl}^{2}$ and $m_{0}^{2}=0.095 M_{pl}^{2}$ \cite{Pawlowski:2023dda}.}
\label{fig05}
\end{figure*}
The mass of the black hole is demonstrated with respect to the event horizon and the $S_{0}$ in Figure (\ref{fig05}). In what follows, this equation can be used to obtain the heat-capacity relationship. Moreover, it is clear that increasing the black hole mass also increases the event horizon, $r_{h}$. 

\begin{figure*}
\centering
{(a) }\includegraphics[width=0.37\linewidth]{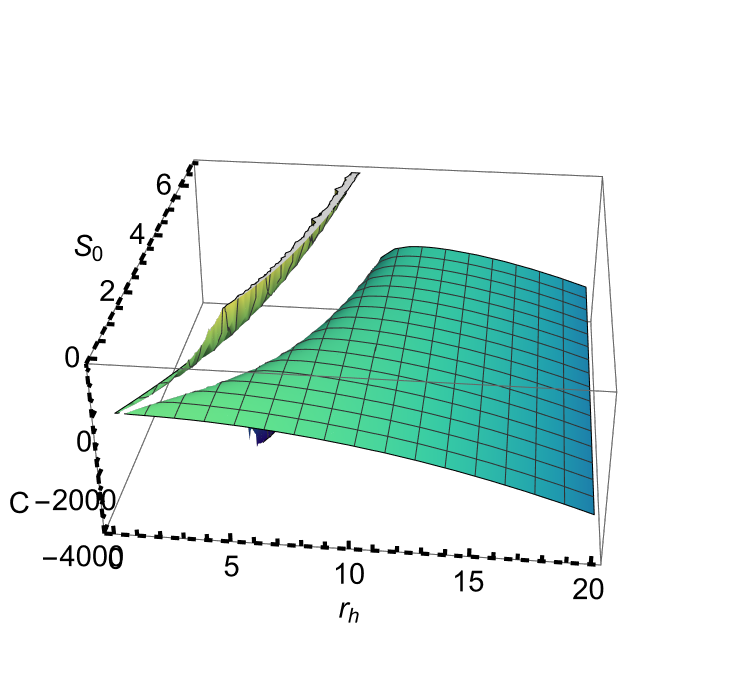}
\includegraphics[width=0.08\linewidth]{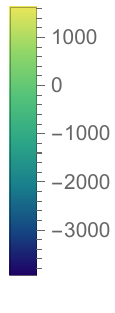}
{(b) }\includegraphics[width=0.26\textwidth]{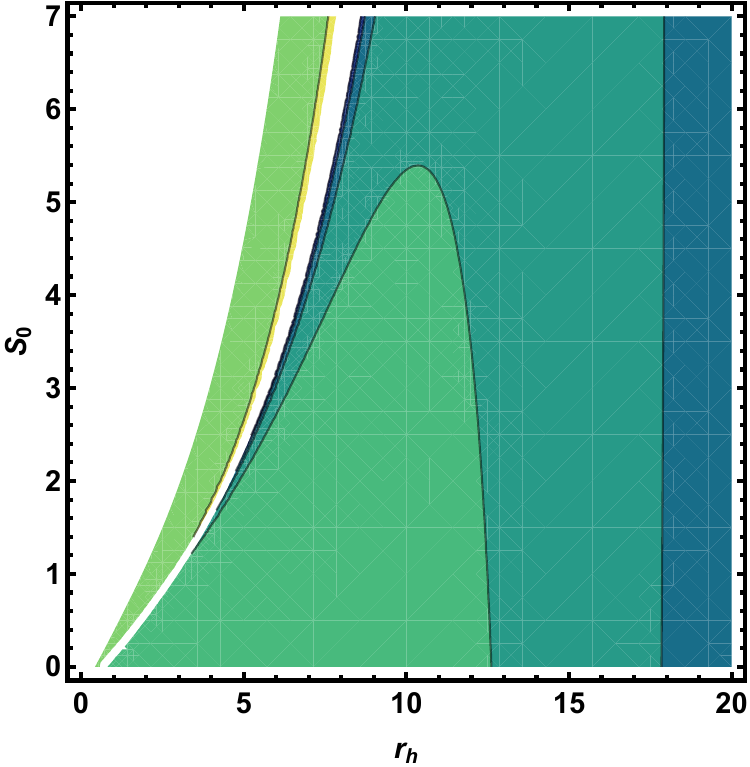}
\includegraphics[width=0.055\linewidth]{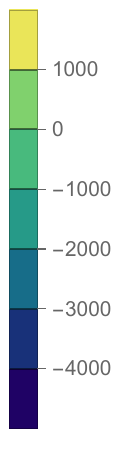}
\caption{The graphs present the possible heat capacity $C$ with respect to variety values of $r_{h}$ and $S_{0}$. We used $m_{2}^{2}=2.5 M_{pl}^{2}$ and $m_{0}^{2}=0.095 M_{pl}^{2}$ \cite{Pawlowski:2023dda}.}
\label{fig06}
\end{figure*}

Another quantity is the heat capacity, and it can be computed from $ C = \frac{dM}{dT}$ \cite{Bonanno:2000ep},
\begin{widetext}
\begin{eqnarray}\label{C}
C = - \Bigg\lbrace 4 e^{- r_{h}(m_{0} + m_{2})} \pi r_{h}^{4} \big( 2 r_{h} m_{0}^{2}S_{0} e^{r_{h}m_{2}} + e^{r_{h}m_{0}} (2 e^{r_{h}m_{2}} - r_{h}m_{2}^{2}S_{2} ) \big) \bigg[ \bigg( e^{- 2 r_{h} (m_{0}+ m_{2})} \big( 2 e^{r_{h}m_{2}}r_{h}m_{0}^{2}S_{0} \nonumber\\ + e^{r_{h}m_{0}} (2 e^{r_{h}m_{2}} - r_{h} m_{2}^{2} S_{2}) \big) \big( -4 S_{0} e^{r_{h}m_{2}} (1 + r_{h}m_{0}) + e^{r_{h}m_{0}} ( 2 r_{h}e^{r_{h}m_{2}} - (1 + r_{h}m_{2}) S_{2} ) \big) \bigg) r_{h}^{-3} \bigg]^{\frac{1}{2}} \nonumber\\  \times \bigg( 8 r_{h} e^{2 r_{h}m_{2}} - 8 r_{h} m_{0}^{2}S_{0}^{2} e^{2 r_{h} (- m_{0} + m_{2})} \big( 2 + r_{h} m_{0} ( 3 + 2 r_{h} m_{0}) \big) - 2 S_{2} e^{r_{h}m_{2}} \big( 3 + r_{h} m_{2} ( 3 \nonumber\\ + r_{h}m_{2} (2 + r_{h}m_{2}) ) \big) + r_{h} m_{2}^{2} S_{2}^{2} \big( 2 + r_{h} m_{2} (3 + 2 r_{h} m_{2}) \big) + 2 S_{0} e^{r_{h}(-m_{0}+m_{2})} \bigg[  2 e^{r_{h}m_{2}} \big( -6 \nonumber\\ + r_{h}m_{0}(-3 +r_{h}m_{0}) (2 + r_{h}m_{0}) \big)  + r_{h} S_{2} \big( - m_{0}^{2} (2 + r_{h} m_{0}) - r_{h} m_{0}^{2} (2 + r_{h}m_{0}) m_{2} \nonumber\\ + ( 2 + r_{h} m_{0})^{2} m_{2}^{2} + 2 r_{h} ( 1 + r_{h} m_{0} ) m_{0}^{3} \big)  \bigg] \bigg)^{-1} \Bigg\rbrace.
\end{eqnarray}
\end{widetext}
One can therefore focus on the heat capacity. According to the Figure (\ref{fig06}), the heat capacity includes the root point related to the phase transition. When the heat capacity is positive, the black hole would be in a stable phase. Also, when it is negative, the black hole would be in an unstable phase. Meanwhile, the heat capacity increases sharply by increasing the value of the event horizon $r_{h}$. It should be mentioned that $C$ bifurcation at a vertical asymptote, separating stable ($C>0$) and unstable ($C<0$) phases.

Another quantity is the entropy, which is obtained from $ dS = \frac{dM}{T} $,

\begin{eqnarray}\label{entropy7}
S=\pi  r_{h}^{2}.
\end{eqnarray}

\begin{figure*}
\centering
{(a) }\includegraphics[width=0.37\linewidth]{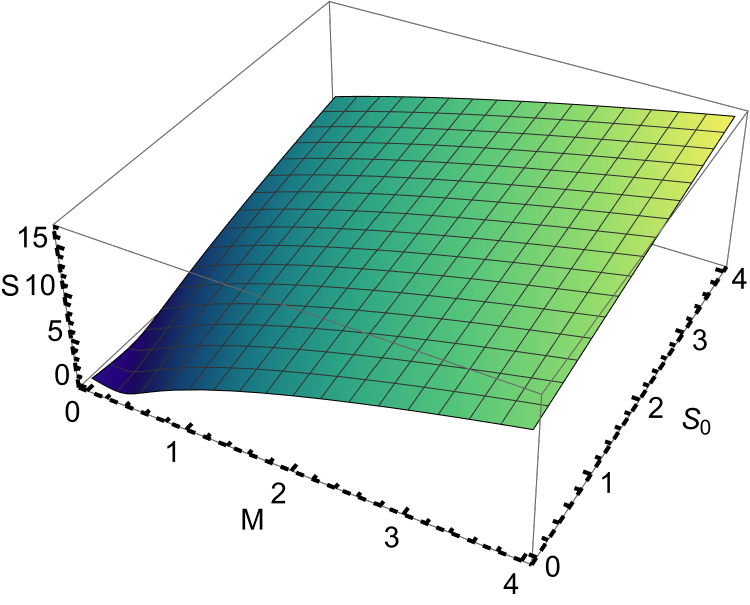}
\includegraphics[width=0.08\linewidth]{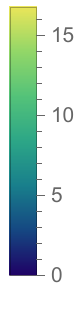}
{(b) }\includegraphics[width=0.26\textwidth]{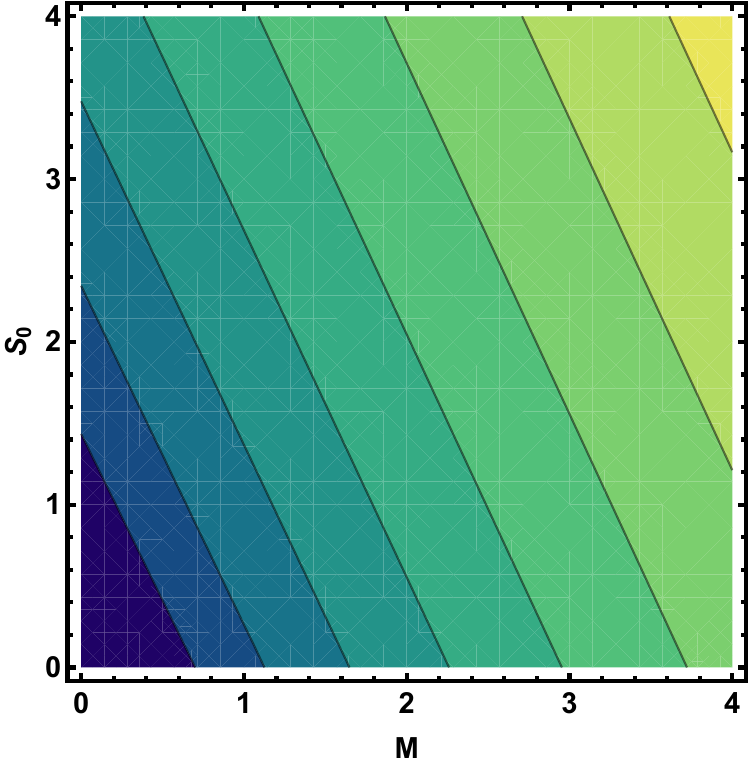}
\includegraphics[width=0.055\linewidth]{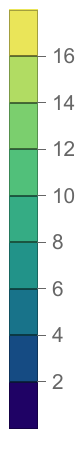}
\caption{The graphs present the possible entropy $S$ with respect to variety values of $S_{0}$ and $M$. We used $m_{2}^{2}=2.5 M_{pl}^{2}$ and $m_{0}^{2}=0.095 M_{pl}^{2}$ \cite{Pawlowski:2023dda}.}
\label{fig07}
\end{figure*}

Figure (\ref{fig07}) illustrates that mass $M$ and parameter $S_{0}$ are competing contributors to the entropy of the black hole. The black hole mass supports the expected thermodynamic view that entropy increases with mass; also, as $S_{0}$ increases, the entropy increases for the same mass. Together, these results help constrain the remaining parameter space for asymptotically safe black holes and black-hole evaporation end states.

\section{ENTANGLEMENT ENTROPY}\label{sec:3}

In this section, we aim to compute the entanglement entropy of Hawking radiation in two cases: without the island and with the island.
Generally, one should present the Kruskal coordinates transformation to analyze the geometry and better understand the calculations. 
Thus, the definition of the tortoise coordinate can be written as,
\begin{equation}
r_{*}= \int \frac{d r}{\sqrt{f(r) g(r)}},
\end{equation}
Also the Kruskal coordinates can be defined as
\begin{equation}
U = - e^{-ku}, \qquad V = e^{kv},
\end{equation}
where
\begin{equation}
u = t - r_{*}, \qquad v = t + r_{*},
\end{equation}
Here $k$ is the surface gravity at the event horizon, which can be obtained,
\begin{equation}
k = \frac{1}{2} \sqrt{f^{'}(r_{h})g^{'}(r_{h})}.
\end{equation}
So, the metric of the black hole in asymptotically safe quantum gravity in the Kruskal coordinates can be written as,
\begin{equation}
d s^{2} = - \Omega^{2} (r) dU dV,
\end{equation}
where the conformal factor $\Omega^{2}(r)$ is given by,
\begin{equation}
\Omega^{2}(r) = \frac{\sqrt{f(r) g(r)}}{k^{2} e^{2 k r_{*}}}.
\end{equation}

\subsection{Without island}\label{subsec41}
 
 Note that we present the entanglement entropy without an island in the early stages of the process. Because of the divergence of entropy in higher dimensions and the lack of an exact formula, we apply the large distance limit to justify the s-wave approximation, which means one neglects the spherical components of the metric and deals mainly with the conformally flat part. Hence, this approach enables a reasonably accurate description of Hawking radiation as perceived by a distant observer. 
Moreover, we restrict our investigation to neutral, non-electrically charged radiation emitted by black holes. Thus, the exclusion of the Schwinger effect is evident from the current scope of this investigation.  

\begin{figure}
\includegraphics[width=0.7\linewidth]{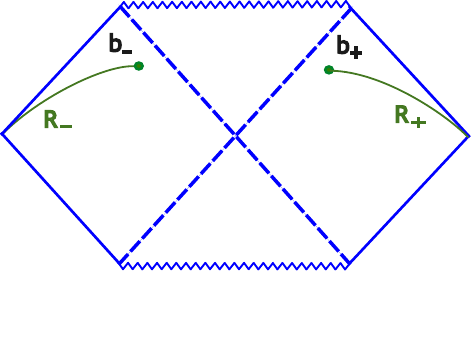}
\caption{This is a Penrose diagram of a black hole in an asymptotically safe quantum gravity background, without an entanglement island. Region $R$, for the Hawking radiation-related degrees of freedom, divides into two parts: $R_{+}$ and $R_{-}$, in the right and left wedges, respectively. The boundary surfaces of $R_{+}$ and $R_{-}$ are denoted by $b_{+}$ and $b_{-}$, respectively.}
\label{fig08}
\end{figure}

According to the Figure (\ref{fig08}), when one considers the absence of the island, there are solely two boundary points of the entanglement regions. Therefore, we can consider the coordinates of these two points as $(t, r) = (t_{b}, b)$ for $b_{+}$ and $(t, r) = (-t_{b} + \frac{i \beta}{2}, b)$ for $b_{-}$. So, the mutual information between the radiation at the left and right wedges can be obtained by \cite{Casini:2005rm},
\begin{equation}
	S_{Bulk}(R) = \frac{c}{3}\log d(b_{+}, b_{-}),
\end{equation}
Here $c$ is the central charge. Moreover, the $d (b_{+}, b_{-})$ is the geodesic distance between the points $b_{+}$ and $b_{-}$. As a result, the geodesic distance can be obtained by
\begin{align}
d(b_{+}, b_{-})^{2} & =  - ds^{2} & \nonumber\\ & =  \Omega^{2}(b) [U(b_{-}) - U(b_{+})][V(b_{+}) - V(b_{-})]. &
\end{align}
By substituting the coordinate of $b_{\pm}$, the entanglement entropy without island is obtained by,
\begin{align}
S_{Bulk}(R) = \frac{c}{6} \log [\frac{4 \cosh^{2} (k t_{b})}{k^{2}}].
\end{align}
In the case of late times limit $t_{b} \rightarrow \infty$, one can use the approximation $\cosh^{2} (k t_{b}) \simeq \frac{1}{4} e^{2 k t_{b}}$, so we have,
\begin{equation}\label{ETE}
S_{Bulk}(R) \simeq \frac{c}{6} \log [\frac{e^{2 k t_{b}}}{k^{2}}] \simeq \frac{c}{3} k t_{b}.
\end{equation}
The entanglement entropy of radiation presents linear growth in time $t_{b}$. Furthermore, because the system is in a pure state at $t=0$, the fine-grained entropy of the radiation will ultimately exceed the Bekenstein-Hawking entropy (a coarse-grained measure) of the black hole in the future. This means that the black hole eventually evolves into a mixed state, providing evidence for a possible unitarity violation. In the absence of an $S$ matrix that ensures unitary evaporation, we encounter the loss of information from the black hole, a phenomenon known as the black hole information paradox.
Theoretical frameworks suggest that introducing the island concept will address this inconsistency and preserve the unitarity principle.  
\\
 
\subsection{With island}\label{subsec42}

\begin{figure}
\includegraphics[width=0.7\linewidth]{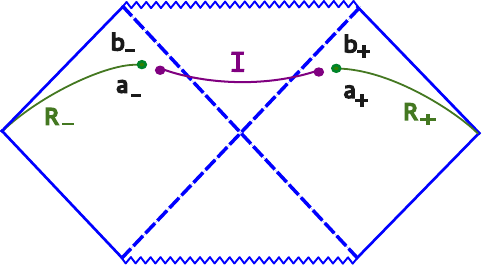}
\caption{This is a Penrose diagram of a black hole in an asymptotically safe quantum gravity background, with an entanglement island. Region $R$, for the Hawking radiation-related degrees of freedom, divides into two parts: $R_{+}$ and $R_{-}$, in the right and left wedges, respectively. The boundary surfaces of $R_{+}$ and $R_{-}$ are denoted by $b_{+}$ and $b_{-}$, respectively. The boundaries of the island, denoted as $I$, are marked by the points $a_{+}$ and $a_{-}$, which are located on the respective edges of the wedges.}
\label{fig09}
\end{figure}

In this subsection, we present the calculation of the entanglement entropy with a single island at the late step of black hole evaporation.
In the two-dimensional case with conformal symmetry, one can obtain the fine-grained entropy formula using renormalization techniques \cite{Holzhey:1994we, Casini:2005rm,Bianchi:2014qua}. Consequently, the entanglement entropy for this disconnected interval is given by \cite{Hashimoto:2020cas},
\begin{equation}
	S_{Bulk}(R \cup I) = \frac{c}{3} \log \frac{d(a_{+},a_{-})d(b_{+},b_{-})d(a_{+},b_{+})d(a_{-},b_{-})}{d(a_{+},b_{-})d(a_{-},b_{+})},
\end{equation}
where $d(x,y)$ is the geodesic distance. It should be noted that the boundaries of the island $I$ are located at $a_{\pm}$ which means $(t,r) = (t_{a}, a)$ for $a_{+}$ and $(t, r) = (-t_{a} + i \frac{\beta}{2}, a)$ for $a_{-}$ (it can be seen in the Figure (\ref{fig09})). Therefore, by considering the Kruskal coordinates and the conformal factor, one can obtain the total entanglement entropy as
\begin{widetext}
\begin{align}
	S_{total} =  \frac{2 \pi a^{2}}{G_{N}} + \frac{c}{6} \log \big[ \frac{16 \sqrt{f(a) g(a)} \sqrt{f(b) g(b)}}{k^{4}} \cosh^{2} (k t_{a}) \cosh^{2}(k t_{b}) \big] \nonumber\\  + \frac{c}{3} \log \big[ \frac{\cosh[k (r_{*}(a) - r_{*}(b))] - \cosh[k (t_{a} - t_{b})]}{\cosh[k(r_{*}(a) - r_{*}(b))] + \cosh[k (t_{a} + t_{b})]} \big].
\end{align}
\end{widetext}

In the following, we analyze the entanglement entropy at early and late times.
We also consider the cutoff surface, located far from the horizon, with $b \gg r_{h}$. 
Early ($t_{a}, t_{b} \approx 0$), one can assume the following approximation as below,
\begin{equation}
t_{a}, t_{b} \approx 0 \quad t_{a}, t_{b} \ll r_{h}, \quad t_{a},t_{b} \ll \frac{1}{k} \ll r_{*}(b) - r_{*}(a). 
\end{equation}
Hence, we can disregard the last term of the total entanglement entropy. Thus, we have the total entropy approximately:
\begin{align}
&S_{total}(early) \simeq \frac{2 \pi a^{2}}{G_{N}}& \nonumber\\& + \frac{c}{6} \log \big[ \frac{16 \sqrt{f(a) g(a)} \sqrt{f(b) g(b)}}{k^{4}} \cosh^{2} (k t_{a}) \cosh^{2}(k t_{b}) \big] & \nonumber\\ & \simeq  \frac{2 \pi a^{2}}{G_{N}} + \frac{c}{6} \big[ \log (\sqrt{f(a) g(a)} \sqrt{f(b) g(b)}) + \cosh (2 k t_{a}) \big]. &
\end{align}
The location of the island can be obtained by solving the following extremizing equations,
\begin{align}
&\frac{\partial S_{total}(early)}{\partial t} = \frac{c k}{3} \sinh (2 k t_{a}) =0, & \nonumber\\ & \frac{\partial S_{total}(early)}{\partial a} = \frac{1}{12} (\frac{48 a \pi}{G_{N}}+\frac{c f^{'}(a)}{f(a)}+\frac{c g^{'}(a)}{g(a)})=0. &
\end{align}
The solution to the first equation yields $ t=0$. In addressing the second equation, using approximation techniques and assuming that higher-order terms of $\frac{a -  r_{h}}{r_{h}}$ are negligible, we come up with a reasonably effective solution (i.e., $a \simeq r_{h}$) along with higher-order terms, which are, by virtue of our assumptions, smaller than the Planck length $l_{p}$. The result suggests that there's no quantum extremal surface that can add entropy to the extremum before Page time. Early on, the entanglement entropy of the radiation is computed without an island contribution. This is because any potential island at early times would have a size on the order of the Planck length or smaller—a scale at which semi-classical gravity breaks down and the island prescription is not applicable. Thus, during the early phase, the entanglement entropy is dominated by radiation alone.
On the other hand, the calculations used to determine the properties of the island assume that it is on a much larger scale, so that the effects of quantum gravity can be washed out; these systems operate at the Planck length. This is because of the notion named "replica symmetry," which is a sort of mathematical symmetry that dissolves when the effect of quantum gravity becomes essential, as shown in studies \cite{Almheiri:2019qdq,Penington:2019kki}. Hence, it's virtually certain that the island is ruled out during this initial phase. The entanglement entropy (which measures quantum information) must be computed during this period without an island in the geometry arrangements. Over time, this entropy increases.

We can now examine the second scenario, the late-time regime, in which the island appears. 
In this case, one should consider the following approximate equations, 
\begin{align}
&\frac{1}{k} \ll r_{*}(b) - r_{*}(a) \ll t_{a},t_{b},& \nonumber\\
&\cosh k (t_{a} + t_{b}) \gg \cosh k (r_{*}(b) - r_{*}(a)), & \nonumber\\
&\cosh k (r_{*}(b) - r_{*}(a)) \simeq \frac{1}{2} e^{k (r_{*}(b) - r_{*}(a))}. &
\end{align}
Furthermore, the right and left wedges are separated by a large distance. So, the separation between points $a_{\pm}$ and $b_{\pm}$ will exhibit the following behavior \cite{Hashimoto:2020cas}:
\begin{equation}
d(b_{+}, b_{-}) \simeq d(a_{+}, a
_{-}) \simeq d(b_{\pm}, a_{\mp}) \gg d(b_{\pm}, a_{\pm}).
\end{equation}
By considering the above conditions, the total entanglement entropy in the late-time can be approximated as below,
\begin{widetext}
\begin{align}
S_{total}(late)   \simeq   \frac{2\pi a^{2}}{G_{N}} + \frac{c}{3} \log \bigg[ \frac{2 \sqrt{f(a) g(a)} \sqrt{f(b) g(b)}}{k^{2}} \big( \cosh \big[ k(r_{*}(a) - r_{*}(b)) \big] - \cosh \big[ k(t_{a}- t_{b}) \big] \big) \bigg]. 
\end{align}
\end{widetext}
Now, we find the derivative of the expression with respect to $t$ and set it to zero,
\begin{equation}
\frac{\partial S_{total}(late)}{\partial t_{a}} = \frac{c}{3} \frac{k \sinh k (t_{a} - t_{b})}{\cosh [k (r_{*}(a) - r_{*}(b))] - \cosh [ k (t_{a} - t_{b})]} = 0.
\end{equation}
Note that the equation can be equal to zero only when $t_{a}$ is approximately equal to $t_{b}$, which implies that the entropy is independent of time. Thus, after substituting the relation and imposing the approximation, the total entanglement entropy in the late-time can be rewritten,
\begin{widetext}
\begin{align}
S_{total}(late)   \simeq  \frac{2\pi a^{2}}{G_{N}} + \frac{c}{3}  \log \bigg[ \frac{ 2 \sqrt{f(a) g(a)} \sqrt{f(b) g(b)}}{k^{2}} \bigg] + \frac{c}{3} k \big( b - r_{*}(a) \big) - \frac{2 c}{3} e^{-k \big( b - r_{*}(a)  \big)}. 
\end{align}
\end{widetext}
Subsequently, we derive the above equation with respect to $a$.
\begin{align}
\frac{\partial S_{total}(late)}{\partial a} = 2 c \big( 1 + 2 e^{k (r_{*}(a) - b)} \big) k r^{'}_{*}(a) \nonumber\\ - \frac{24 a \pi}{G_{N}} - c \big[ \frac{f^{'}(a)}{f(a)} + \frac{g^{'}(a)}{g(a)} \big] = 0,
\end{align}
one can expand $f(r)$ and $g(r)$ linearly around $r_{h}$
\begin{align}
&f(r) \simeq f^{'}(r_{h}) (r - r_{h}), &\nonumber\\ & g(r) \simeq g^{'}(r_{h})(r - r_{h}). &
\end{align}
At this stage, the tortoise coordinate can be approximated as follows. For consistency with the definition of $k$, assume the quadratic term dominates. So, we have
\begin{align}
r_{*}(r) = \int \frac{1}{\sqrt{f(r) g(r)}} dr \simeq  \int \frac{dr}{2k (r- r_{h})} = \frac{1}{2 k} \ln |\frac{r - r_{h}}{r_{h}}|.
\end{align}
It is worth pointing out that one can consider the situation the island would be near the event horizon which means $a \simeq r_{h} + \epsilon$ with $(\epsilon^{2} = \frac{c^{2}G_{N}^{2}}{576 \pi^{2} r_{h}^{2}}  \ll 1)$.
After substituting and imposing expansion, the entanglement entropy can be obtained.
\begin{equation}\label{ETL}
S_{total}(late) = \frac{2 \pi r_{h}^{2}}{G_{N}} + \mathcal{O}(1) \simeq 2 S_{BH},
\end{equation}

The $\mathcal{O}(1)$ corrections include constant terms, $\mathcal{O}(\epsilon^{2})$, and a negative logarithmic contribution $\frac{c}{6}\log |\epsilon|$, but these are negligible compared to the leading area term for macroscopic black holes.

It should be noted that the additional parameters of the black hole contribute directly and indirectly to the position of island boundaries through altering the radius of the event horizon $r_h$.

Regarding late-time entanglement entropy, the second term is finite. Thus, in late times, the Bekenstein-Hawking entropy contributes most to entanglement entropy. Black hole entanglement entropy is time-independent throughout black hole evaporation, such that the final state is unitary. Page curve compatibility supports information recovery and unitary Hawking radiation. In this case, the black hole's degrees of freedom, which can entangle, effectively lower the black hole's fine-grained entropy, putting it below the coarse-grained Bekenstein-Hawking limit.

The island boundary $a$ is a function of $r_{h}$ , and so is sensitive to $S_{0}$ , $m_{0}$, and $m_{2}$. Additionally, an asymptotically safe quantum gravity black hole has parametrically tunable island locations, unlike an ordinary Schwarzschild black hole. Here, the decreases in $r_{h}$ shift the islands closer to the horizon, thus increasing the contribution of the quantum extremal surfaces.

\subsection{Page time and scrambling time}\label{subsec43}

In this subsection, we present the Page time and scrambling time using entanglement-entropy values at early and late times. Hence, we equate the equations (\ref{ETE}) and (\ref{ETL}) to obtain the Page time.
\begin{equation}\label{tPage}
t_{Page} \simeq \frac{6 S_{BH}}{ck} \simeq \frac{3 S_{BH}}{c \pi T}.
\end{equation}

We therefore focus on the scrambling time. Scrambling time is defined as the earliest time at which an outside observer must recover quantum information from Hawking radiation before a black hole absorbs it. The idea is based on the Hayden-Preskill thought experiment. This theoretical model posits that, after the Page time (the time at which a black hole begins to lose mass), one must wait until the scrambling time to reconstruct information from previously emitted radiation \cite{Hayden:2007cs}. At this time, quantum data is expected to arrive at the edge of the island of entanglement. The equation below describes the time it takes light to travel from a point on the island to its boundary.
\begin{align}
& t_{scr} \equiv Min [\Delta t] = r_{*}(b) - r_{*}(a) & \nonumber\\ & = \frac{1}{k} \log \bigg( \frac{12\sqrt{\pi r_{h}^{3}(b - r_{h})}e^{b k}}{c G_{N}} \bigg) &\nonumber\\ & \simeq \frac{1}{k} \log S_{BH} \simeq \frac{1}{2 \pi T} \log S_{BH}, &
\end{align}
The scrambling time in this work is proportional to the logarithm of the Bekenstein-Hawking entropy, which is in accord with the fast scrambler conjecture \cite{Sekino:2008he}. This shows that the decoding associated with the scrambling time can be effectively explained through the Hayden-Preskill protocol \cite{Hayden:2007cs}. We computed the entanglement entropy by locating the boundary of the entanglement island and using it, showing a striking correspondence with the expected Page curve of unitary black holes. Additionally, using the island prescription, we can derive the scrambling time, which provides deeper insight into the dynamics involved and supports the framework we've proposed. Those findings play their respective roles in the unfolding story of quantum information and black hole thermodynamics.

The logarithmic scaling, $t_{scr}$, is consistent with the fast-scrambling conjecture, which shows that black holes are the absolute fastest information processors. Interestingly, this scaling persists even after the UV modifications of asymptotically safe quantum gravity, suggesting a universal behavior in quantum information recovery.

\subsection{Comparison between thermodynamic phase transition and Page time transition}\label{subsec44}

Two critical phases in the life cycle of an evaporating black hole are addressed in the present study. The thermodynamic phase transition at the critical mass $M_{crit}$ is related to the thermodynamic stability of the black hole and heat capacity. Moreover, the Page time transition $t_{Page}$ indicates the transition of information. Whether these events occur in chronological order remains an open question.

To analyze and determine which occurs first, we should calculate and compare the times of the thermodynamic phase transition and the Page-time transition. 
Therefore, we first present the value of Page time using equation (\ref{tPage}) by considering the obtained equations (\ref{rh}), (\ref{T}), and (\ref{entropy7}).

\begin{align}
t_{Page}(M_{i}) = \frac{3 r_{h}^{2}}{c T(r_{h})}.
\end{align}

Note that we can treat the initial mass $M_{i}$ as physically meaningful when calculating the Page time. So, one can consider initial mass of black hole $M_{i}= 10 - 10^{7} M_{pl}$, and a choice of parameters $S_{0}=2$, $m_{2}^{2}=2.5 M_{pl}^{2}$, and $m_{0}^{2}=0.095 M_{pl}^{2}$, we can compute $t_{Page}(M_{i})$ as we show as below table. 

We now aim to determine the time of the thermodynamic phase transition. Note that it is a moment in the black hole's evolution, defined by the specific event horizon $r_{crit}$ and mass $M_{crit}$. According to the Figure (\ref{fig06}), this is the event horizon at which the heat capacity $C$ bifurcates. Thus, when the heat capacity is zero or changes sign, we can find $r_{crit}$ by substituting it into equation (\ref{M}), and then obtain $m_{crit}$.

Now we need to find the time it takes for the black hole to evaporate from its initial mass $M_{i}$ down to the critical mass $M_{crit}$.
It is pointed out that the mass loss rate is given by the Stefan-Boltzmann law for Hawking radiation. The mass loss rate due to Hawking evaporation follows from treating the black hole as a blackbody radiator \cite{Frolov:1998wf}. For a black hole with horizon area $A$ and temperature $T$, the power radiated is given by,

\begin{align}
\frac{dM}{dt} \sim - \sigma A T^{4},
\end{align}

Where $\sigma$ incorporates the Stefan-Boltzmann constant and the number of relativistic degrees of freedom contributing to the radiation \cite{Frolov:1998wf,Page:1976df,Page:2004xp}. In our asymptotically safe gravity framework, both $A$ and $T$ are modified according to Eqs. (\ref{rh}) and (\ref{T}), but the fundamental blackbody relation remains valid. Therefore, the time evaporates from $M_{i}$ to $M_{crit}$ is obtained by,

\begin{align}
t_{crit} \sim \int_{M_{crit}}^{M_{i}} \frac{dM}{\sigma 4 \pi r_{h}^{2}T^{4}}.
\end{align}

In this stage, we present the times of Page phase transition and thermodynamic phase transition in the below table by considering the parameters $S_{0}=2$, $m_{2}^{2}=2.5 M_{pl}^{2}$, and $m_{0}^{2}=0.095 M_{pl}^{2}$.

\begin{table*}
\caption{\label{tab:table1} The table presents the times of Page phase transition and thermodynamic phase transition by considering the parameters $M_{crit}= 1.47 M_{pl}$, $S_{0}=2$, $m_{2}^{2}=2.5 M_{pl}^{2}$, and $m_{0}^{2}=0.095 M_{pl}^{2}$.}
\scalebox{1}{
\begin{tabular}{|c|c|c|c|}\hline
$M_{i}$ $(M_{pl})$ &$t_{Page}$ & $t_{crit}$ & $\frac{t_{crit}}{t_{Page}}$ \\ \hline
$10^{1}$ & $1.706$ & $1.493 \times 10^{9}$ & $ >> 1$ \\ \hline
$10^{2}$ & $1.735$ & $2.265 \times 10^{9}$ & $ >> 1$ \\ \hline
$10^{3}$ & $2.064$ & $2.350 \times 10^{9}$ &  $>> 1$ \\ \hline 
$10^{4}$ & $2.771$ & $2.358 \times 10^{9}$ &  $>> 1$ \\ \hline 
$10^{5}$ & $3.912$ & $2.359 \times 10^{9}$ &  $ >> 1$ \\ \hline 
$10^{6}$ & $5.642$ & $2.359 \times 10^{9}$ &  $ >> 1$ \\ \hline 
$10^{7}$ & $8.216$ & $2.359 \times 10^{9}$ &  $ >> 1$ \\ \hline 

\end{tabular}}
\end{table*}

\begin{align}
t_{crit} >> t_{Page}.
\end{align}

Our analysis reveals that the thermodynamic phase transition time is much longer than $t_{crit}$, the Page time $t_{Page}$. We consistently find that this issue occurs for all considered values of $S_{0}$ and initial masses $M_{i}$. Thus, this indicates that the Page transition occurs significantly earlier in the evaporation process than the thermodynamic phase transition, which occurs at the change in sign of the heat capacity.

It is worth pointing out that long before the black hole reaches the thermodynamic phase transition and thermodynamic stability changes, the fine-grained entropy of radiation has already saturated at twice the Bekenstein–Hawking entropy. Therefore, this supports the unitary evolution of the black hole, as information is encoded in Hawking radiation well before the black hole enters its final unstable phase.
The observed analysis is across a range of $S_{0}$ values and initial masses, indicating that the early onset of information recovery is a generic feature of asymptotically safe black holes, not fine-tuned to specific parameters.

\section{Conclusion}\label{sec:4}

We have investigated the thermodynamic properties and the entanglement island of a black hole in asymptotically safe quantum gravity. 

First, we derived thermodynamic quantities such as mass $M$, Hawking temperature $T$, and heat capacity $C$, and their dependence on the event-horizon radius $r_{h}$. Hence, we have shown that the event horizon radius is positively correlated with black hole mass and that an increase in the parameter $S_{0}$ corresponds to a rise in the event horizon radius.

Moreover, we have shown the Hawking temperature $T$ as a function of mass $M$ and the parameter $S_{0}$. The evaporation process also indicates that black holes lose mass via Hawking radiation until they reach a phase transition, after which evaporation stops. This shows that there is a natural stopping point from the lowest order of mass loss due to Hawking evaporation.

In addition, we have derived a relationship among the black hole mass, the event-horizon radius, and the temperature to calculate the heat capacity. In this case, the event horizon radius increases with increasing mass and follows the classical picture of gravitational collapse. Meanwhile, lower-mass black holes have lower temperatures, which is inconsistent with classical predictions. 

We have also demonstrated thermodynamic stability using heat capacity, where the point at which the heat capacity sign changes could be interpreted as a phase transition from a stable (positive heat capacity) to an unstable (negative heat capacity) state. The heat capacity exhibits interesting critical behavior, bifurcating sharply as a function of $r_{h}$.
In the Schwarzschild black hole, the heat capacity is negative, and the black hole is unstable; thus, as it evaporates and its mass decreases, the temperature increases. However, in our study, the observed thermal gradient in the black hole spectrum is inconsistent with classical predictions, implying that decreasing the black hole mass also reduces the temperature. This point is completely consistent with the observed behaviour of the heat capacity of this black hole. 

Therefore, we indicated that the entropy of the black hole rises by increasing the mass of the black hole $M$, and it grows gradually by increasing the value of the $S_{0}$. 

Subsequently, we have studied the entanglement entropy of Hawking radiation both with and without island contributions.
In the case of a non-island, the entanglement entropy of radiation exhibits linear growth with time $t_{b}$. Since the system begins in a pure state at $t=0$, the fine-grained entropy of radiation eventually surpasses the Bekenstein-Hawking (coarse-grained) entropy of the black hole at late times. Without corrective mechanisms, this would imply a transition to a mixed state, violating unitarity and giving rise to the black hole information paradox due to the absence of a unitary S-matrix for evaporation. 

Introducing entanglement islands resolves the problem of unbounded entropy production, thereby ensuring unitarity. The boundaries of the entanglement islands and their locations also depend on the black hole's properties. This occurs by altering the horizon radius $r_{h}$. At late times, the entanglement entropy is dictated by the Bekenstein-Hawking term, rendering the entropy finite and time-independent. This asymptotic behavior is in agreement with the Page curve. It indicates that (i) there is no problem with Hawking radiation being unitary, (ii) information does not go lost, and (iii) the fine-grained entropy of the black hole is less than the coarse-grained entropy, as the degrees of freedom of the black hole entangle entirely with the degrees of freedom of the radiation.

The calculations we performed using the island prescription explicitly matched expectations of the unitary Page curve. And, crucially, because we also calculated the scrambling time, we provide insight into quantum information, consistent with Page-curve expectations. Taken together, these outcomes represent a significant advance in resolving the problem of black hole thermodynamics within quantum mechanics.

The findings presented here suggest that the black hole in asymptotically safe quantum gravity is naturally consistent with unitary evaporation and the recovery of information via entanglement islands. Future studies should include charged and/or rotating solutions, as well as fluctuations of the renormalization-group scale within the asymptotic safety framework, which could yield testable predictions, such as gravitational-wave signals or signatures in remnant decay.

\section*{Acknowledgements}
This work is supported by the National Natural Science Foundation of China (No. 12105013).
 

\end{document}